\documentclass[showpacs,preprintnumbers,prd,nofootinbib,floats,amssymb,floatfix]{revtex4}
\usepackage{graphicx}
\usepackage{amsxtra}
\usepackage{amssymb}
\usepackage{amsmath}
\usepackage{bbold}
\usepackage{physics}
\usepackage{bm}
\usepackage{tensor}
\usepackage{xcolor}
\usepackage{graphicx}
\graphicspath{ {Imagenes/} }
\usepackage{hyperref}
\usepackage{amssymb}
\usepackage{amstext}
\usepackage{amsmath}
\usepackage{cleveref}
\usepackage{stackrel}
\usepackage{blkarray}
\begin{document}
\title{The symplectic  analysis for  Barbero-Husain-Kucha\v{r} gravity }
\author{Diana Vanessa Castro-Luna}  \email{dcastro@ifuap.buap.mx}
\author{Alberto Escalante} \email{aescalan@ifuap.buap.mx} 
\affiliation{Instituto de F\'isica, Benem\'erita Universidad Aut\'onoma de Puebla. \\ Apartado Postal J-48 72570, Puebla Pue., M\'exico,}
%\affiliation{Instituto de F\'isica, Benem\'erita Universidad Aut\'onoma de Puebla. \\ Apartado Postal J-48 72570, Puebla Pue., M\'exico,}
%\affiliation{Instituto de F\'isica, Benem\'erita Universidad Aut\'onoma de Puebla. \\ Apartado Postal J-48 72570, Puebla Pue., M\'exico,}
\author{Jos\'e Gustavo Kuri Ric\'ardez }
\affiliation{Facultad de Ciencias Físico Matemáticas, Benemérita Universidad Autónoma de Puebla, Apartado postal
1152, 72001 Puebla, Pue., Mexico}
	\begin{abstract}
		A comprehensive Faddeev-Jackiw symplectic analysis of the Barbero-Husain-Kuchař  $[BHK]$ gravitational action is performed. By isolating the algebraic structure of the canonical momenta $\Pi^a_i$, we demonstrate that the spatial components of the auxiliary vector field $v_\alpha$ are not free Lagrange multipliers ; rather, they encode the three antisymmetric components of the momenta within the triad frame. Through computing the zero modes of the symplectic matrix, we prove that the auxiliary triad sector has only two algebraic gauge symmetries. Consequently, imposing the three conditions $v_a = 0$ by hand as a gauge choice, as previously assumed in the literature to recover General Relativity, is mathematically inconsistent and over-constrains the phase space. To address this problem, we propose a modified action that couples the vector field $v_\alpha$ to a clock scalar field $\phi$ through an antisymmetric tensor field $\Lambda^{\alpha\beta}$. We show that, in the relational clock gauge, the condition $v_a = 0$ emerges dynamically as a genuine constraint derived from the variational principle. This promotes the so-called scalar constraint to a true phase-space constraint, and successfully recovers the two physical degrees of freedom of Einstein's General Relativity.
		\end{abstract}
	
	\date{\today}
	
	
	\preprint{}
	
	\maketitle
\section{Introduction}
	In the quest for a non-perturbative quantization of gravity, background-independent gauge formulations of General Relativity [GR] have played a pivotal role~\cite{Ashtekar1986, Barbero1995, Rovelli2004, Thiemann2007}. Among the most insightful diffeomorphism-invariant models studied in the canonical framework are the BF theories, $(2+1)$-dimensional gravity in its Chern--Simons formulation~\cite{Witten1988}, and the Husain-Kucha\v{r} (HK) model~\cite{HusainKuchar1990}.  The HK theory is  a four-dimensional $SU(2)$ gauge theory based on  a triad one-form $e^i_\alpha$ and an $SU(2)$ connection $A^i_\alpha$ whose curvature is $F^i_{\mu\nu}$. The HK action shares the kinematical phase space of Ashtekar-Barbero gravity and includes both the internal $SU(2)$ Gauss law and the three-dimensional spatial diffeomorphism constraints. However, as it lacks the Hamiltonian (scalar) constraint which generates time evolution normal to the spatial hypersurfaces, HK gravity is a ``timeless" theory that propagates $3$ local physical degrees of freedom per spatial point, rather than the $2$ degrees of freedom of full Einstein gravity.\\
	%%%
	To bridge the gap between the HK model and full General Relativity with a real $SU(2)$ connection, Barbero~\cite{Barbero1994} introduced a modified action the so-called  Barbero-Husain-Kucha\v{r} [BHK] gravity by supplementing the HK Lagrangian with a linear coupling to an external covector field $v_\alpha$ of the form $\epsilon^{\alpha\beta\mu\nu}v_\alpha e_{\beta i}F^i_{\mu\nu}$. The underlying motivation behind this construction is that the temporal component $v_0$ acts as a Lagrange multiplier enforcing the additional scalar constraint $\Omega = \eta^{abc}e_{ai}F^i_{bc} = 0$, which, when combined with the vector constraint, is expected to reduce the $3$ degrees of freedom of the HK model down to the $2$ physical polarizations of GR. Crucially, however, achieving the canonical constraint algebra of GR in this framework has traditionally relied on assuming that the spatial components of the vector field can be fixed to zero, $v_a = 0$, as a gauge condition~\cite{Barbero1994}. Whether such an \textit{a priori} truncation of $v_a$ is dynamically and symplectically permissible is a subtle question that requires a rigorous treatment of the full constraint structure and the zero modes of the theory. While Dirac's canonical method~\cite{Dirac1964} is the standard tool for analyzing singular systems, the Faddeev-Jackiw (FJ) symplectic formalism~\cite{FaddeevJackiw1988, BarcelosNeto1992, Montani1993} provides a geometrically transparent and algebraically economical alternative. By working solely with first-order symplectic Lagrangians and identifying constraints through the null vectors of the  generalized symplectic two-form, the FJ method avoids the artificial proliferation of primary conjugates to non-dynamical multipliers and directly exposes the true gauge symmetries and second-class-kind  sectors of the action~\cite{fad2,fad3,fad4,fad5}.\\
	With all exposed above, in this paper, we perform a detailed FJ symplectic analysis of the Barbero-Husain-Kucha\v{r} action and uncover a fundamental obstruction to setting $v_a = 0$ by hand. The paper is organized as the following. In Section~II, we examine the original BHK action and show that the definition of the canonical momenta $\Pi^a_i$ conjugate to $A^i_a$ mixes the triad $e_{ai}$  and the spatial vector $v_a$. By inverting this relation in the triad frame, we prove two key structural results: The three spatial components $v_d$ are uniquely absorbed into the antisymmetric part of the canonical momenta via the following relation $v_d = \frac{1}{4}\eta_{dga}e^{ig}\Pi^a_i$, whereas eliminating $v_a$ leaves behind six independent symmetric constraints, say,  $\phi^{ab}(\Pi, e) = 0$ that relate the symmetric part of the momenta to the spatial metric $q^{ab} = e^{ia}e^{ib}$.
	When $\phi^{ab} = 0$ and the scalar constraint  are incorporated into the symplectic Lagrangian, they form seven independent algebraic second-class-kind pairs with the nine components of the auxiliary triad $e_{ai}$. Consequently, the symplectic matrix $f^{(1)}_{\alpha\beta}$ possesses a nullity of eight: three generators of $SU(2)$ rotations, three generators of spatial diffeomorphisms, and only \textit{two} algebraic zero modes in the internal triad sector.
	Because one can only impose as many algebraic gauge-fixing conditions as there are algebraic null vectors, attempting to impose the three conditions $v_a = 0$ as a gauge choice over-constrains the system and artificially mutilates the antisymmetric sector of the canonical momenta $\Pi^a_i$. Properly accounting for all symplectic variables reveals that the scalar constraint is consumed in fixing one internal orientation of the auxiliary triad rather than constraining the canonical pair $(A^i_a, \Pi^a_i)$, so the  BHK action still propagates $3$ physical degrees of freedom.\\
	To consistently recover the $2$ physical degrees of freedom of General Relativity, the vanishing of $v_a$ must not be imposed \textit{ad hoc}, but must instead emerge dynamically from the variational principle. In Section~III, we construct an alternative action by coupling the vector field $v_\alpha$ to a scalar field $\phi$ through an antisymmetric tensor multiplier $\Lambda^{\alpha\beta}$ via the term $\Lambda^{\alpha\beta}v_\alpha\partial_\beta\phi$. Applying the Faddeev-Jackiw formalism to this extended system, we demonstrate that the new coupling generates the constraint $\Omega_{ab} = v_a\partial_b\phi - v_b\partial_a\phi = 0$ alongside a modified diffeomorphism and scalar constraint structure. Upon adopting the relational clock gauge, the symplectic equations of motion dynamically enforce $v_a = 0$ and decouple the auxiliary sector entirely. As a result, the scalar constraint  becomes a genuine seventh constraint on the $18$-dimensional canonical phase space $(A^i_a, \Pi^a_i)$, cleanly yielding the $2$ physical degrees of freedom of Einstein's General Relativity.
\section{Symplectic analysis of  Barbero-Husain-Kucha\v{r} gravity }
We start with the Barbero-Husain-Kuchař action 
\begin{equation}
S = \int d^4x \left[ \epsilon^{\alpha\beta\mu\nu} v_\alpha e_{\beta i} F^i_{\mu\nu} - \frac{1}{2} \epsilon^{\alpha\beta\mu\nu} \epsilon_{ijk} e^i_\alpha e^j_\beta F^k_{\mu\nu} \right],
\end{equation}
and we identify the following symplectic Lagrangian 
\begin{equation}
\mathcal{L}^{(0)} = \Pi^a_i F^i_{0a} + \eta^{abc} v_0 e_{ai} F^i_{bc} - \eta^{abc} v_a e_{0i} F^i_{bc} - \eta^{abc} \epsilon_{ijk} e^i_0 e^j_a F^k_{bc},
\end{equation}
where  $\Pi^a_i \equiv \eta^{abc} \left( 2 v_b e_{ci} - \epsilon_{ijk} e^j_b e^k_c \right)$.  The symplectic Lagrangian, can be written as $\mathcal{L}^{(0)} = \Pi^a_i \dot{A}^i_a - V^{(0)}$, where $V^{(0)} = -A^i_0 D_a \Pi^a_i - \eta^{abc} v_0 e_{ai} F^i_{bc} + \eta^{abc} e_{0i} \left( v_a F^i_{bc} + \epsilon^i{}_{jk} e^j_a F^k_{bc} \right)$. Hence, the symplectic Lagrangian has the following form $\mathcal{L}^{(0)} = a^{(0)}_\alpha \dot{\xi}^{(0)\alpha} - V^{(0)}$. We identify the symplectic variables  $\xi^{(0)\alpha} = \left( A^i_0, A^i_a, e_{0i}, e_{ai}, v_0, v_a \right)$ and the one-forms $a^{(0)}_\alpha = \left( 0, \; \eta^{abc}\left( 2 v_b e_{ci} - \epsilon_{imn} e^m_b e^n_c \right), \; 0, \; 0, \; 0, \; 0 \right)$; however, we will perform some changes for studying the system. First, from   the definition of the momenta, we find that $v_d = \frac{1}{4} \eta_{dga} e^{ig} \Pi^a_i$. In this manner, we can use that expression and the symplectic Lagrangian will take the form
\begin{equation}
\mathcal{L}^{(0)} = \Pi^a_i \dot{A}^i_a - V^{(0)},
\end{equation}
where 
\begin{equation}
V^{(0)} = -A^i_0 D_a \Pi^a_i - \eta^{abc} v_0 e_{ai} F^i_{bc} + e_{0i} \left( \frac{1}{2} e^{jb} \Pi^c_j F^i_{bc} + \eta^{abc} \epsilon^i{}_{jk} e^j_a F^k_{bc} \right), 
\end{equation}
 thus, after removing the $v_a's$ variables, the new symplectic variables will be given by $\xi^{(0)\alpha} = \left( A^i_0, \; A^i_a, \; \Pi^a_i, \; e_{0i}, \; e_{ai}, \; v_0 \right)$ and the one-forms are $a^{(0)}_\alpha = \left( 0, \; \Pi^a_i, \; 0, \; 0, \; 0, \; 0 \right)$. At this stage of the analysis, one might assume that eliminating the dependence on $v_a's$ is enough, but that is not the case. In fact, if we use the expression  of $v_a$ in the momenta, we obtain 
 \begin{equation}
  \phi^b_i:\frac{1}{2} \Pi^b_i + \frac{1}{2} e_{ci} e^{jb} \Pi^c_j  + \eta^{bcd} \epsilon_{ijk} e^j_c e^k_d=0.
 \end{equation}
Therefore, by taking the antisymmetric part of $\phi^b_i$, it follows that $\eta_{fga}e^{gi} \phi^a_i \equiv 0$; this implies the existence of $9-3=6$ components, $3$ of which are absorbed by the $v'_{a}s$ variables. On the other hand, for the symmetric part of $\phi^b_i$, if we calculate the projection $\phi^{ab} = \phi^a_i e^{bi}$, we obtain six independent components, which are given by
 \begin{equation}
 \phi^{ab}: \frac{1}{2} e^{bi}\Pi^{a}_{i}+  \frac{1}{2}e^{ai}\Pi^{b}_{i} + \eta^{agc} \epsilon_{ijk} e^{j}_g e^{k}_c e^{bi}=0,
 \end{equation}
 here, we note that $\eta^{agc} \epsilon_{ijk} e^{j}_g e^{k}_c e^{bi} = 2eq^{ab}$, where $q^{ab} = e^{ia} e^{ib}$ denotes the spatial metric. Throughout this article, we will work with the triad. Therefore, the six independent components of $\phi^{ab}$ constitute the constraints of the theory and must be added to the symplectic Lagrangian. Thus, the symplectic Lagrangian takes the form
 \begin{equation}
\mathcal{L}^{(0)} = \Pi^a_i \dot{A}^i_a - V^{(0)},
\end{equation}
where 
\begin{equation}
V^{(0)} = -A^i_0 D_a \Pi^a_i - \eta^{abc} v_0 e_{ai} F^i_{bc} + e_{0i} \left( \frac{1}{2} e^{jb} \Pi^c_j F^i_{bc} + \eta^{abc} \epsilon^i{}_{jk} e^j_a F^k_{bc} \right) + \lambda_{ab}\phi^{ab}, 
\end{equation}
we can identify the set symplectic variables  $\xi^{(0)\alpha} = \left( A^i_0, \; A^i_a, \; \Pi^a_i, \; e_{0i}, \; e_{ai}, \; v_0, \lambda_{ab} \right)$ and the one-forms $a^{(0)}_\alpha = \left( 0, \; \Pi^a_i, \; 0, \; 0, \; 0, \; 0, 0 \right)$. From the symplectic variables we obtain the following symplectic matrix 
\begin{equation}
f^{(0)}_{\alpha\beta}(x,y) = \frac{\delta a^{(0)}_\beta(y)}{\delta \xi^{(0)\alpha}(x)} - \frac{\delta a^{(0)}_\alpha(x)}{\delta \xi^{(0)\beta}(y)}, 
\end{equation}
where 
\begin{equation}
f^{(0)}_{\alpha\beta}(x,y) = 
\begin{pmatrix} 
	0 & 0 & 0 & 0 & 0 & 0 & 0 \\ 
	0 & 0 & -\delta^b_a \delta^i_j & 0 & 0 & 0 & 0 \\ 
	0 & \delta^a_b \delta^j_i & 0 & 0 & 0 & 0 & 0 \\
	 0 & 0 & 0 & 0 & 0 & 0 & 0 \\ 
	 0 & 0 & 0 & 0 & 0 & 0 & 0 \\ 
	 0 & 0 & 0 & 0 & 0 & 0 & 0 \\ 
	 0 & 0 & 0 & 0 & 0 & 0 & 0 
	 \end{pmatrix} \delta^3(x-y),
\end{equation} 
and the symplectic matrix has the following null vectors
\begin{eqnarray}
\mathcal{V}^{(0)1}_\alpha& =& \left( \delta^i_k, \; 0, \; 0, \; 0, \; 0, \; 0, \; 0 \right), \nonumber \\
\mathcal{V}^{(0)2}_\alpha &=& \left( 0, \; 0, \; 0, \; \delta^k_i, \; 0, \; 0, \; 0 \right), \nonumber \\
\mathcal{V}^{(0)3}_\alpha &=& \left( 0, \; 0, \; 0, \; 0, \; \delta^a_d \delta^i_k, \; 0, \; 0 \right), \nonumber \\
\mathcal{V}^{(0)4}_\alpha &=& \left( 0, \; 0, \; 0, \; 0, \; 0, \; 1, \; 0 \right), \nonumber \\
\mathcal{V}^{(0)5}_\alpha &=& \left( 0, \; 0, \; 0, \; 0, \; 0, \; 0, \; \delta^{(a}_c \delta^{b)}_d \right), 
\end{eqnarray}
hence, form these null vectors, we obtain  by means of 
\begin{equation}
		\label{cons}
		\Omega_k= \int dx^3 \mathcal{V}_k^{\alpha}\frac{\partial}{\partial \xi^{\alpha}(x)}\int V^{(0)}(\xi(y))dy^3=0, 
	\end{equation}
	the following constraints 
\begin{eqnarray}
\Omega_i &\equiv& D_a \Pi^a_i = 0, \nonumber \\
\Omega^i &\equiv& \frac{1}{2} e^{jb} \Pi^c_j F^i_{bc} + \eta^{abc} \epsilon^i{}_{jk} e^j_a F^k_{bc} = 0, \nonumber \\
\Omega &\equiv& \eta^{abc} e_{ai} F^i_{bc} = 0,  \nonumber \\
\phi^{ab} &\equiv& \frac{1}{2} e^{bi} \Pi^a_i + \frac{1}{2} e^{ai} \Pi^b_i + \eta^{agc} \epsilon_{ijk} e^j_g e^k_c e^{bi} = 0, 
\end{eqnarray}
from $\mathcal{V}^{(0)3}_\alpha$ we obtain algebraic relations between the Lagrange multipliers. On the other hand, we can observe that  by using these constraints, we can obtain the following relation
\begin{equation}
\Pi^b_i F^{i}_{ab}= -v_a\Omega+ e_{ai}\Omega^i+ \phi^{b}_iF^i_{ab}, 
\end{equation}
thus, using $e_0^i= N^ae_a^i$, redefining the Lagrange multipliers as $\tilde v_0= v_0 - N^av_a$, and adding the constraints, we can rewrite the symplectic Lagrangian as
\begin{equation}
\mathcal{L}^{(1)} = \Pi^a_i \dot{A}^i_a + \dot{\lambda}^i \Omega_i(A, \Pi) + \dot{\tilde{\mu}}^a \mathcal{H}_a(A, \Pi) + \dot{\tilde{\rho}} \, \Omega(A, e) + \dot{\tilde{\beta_{ab}}} \phi^{ab}- V^{(1)}, 
\end{equation}
where $V^{(1)}=0$, and $\mathcal{H}_a(A, \Pi) = \Pi^b_i F^{i}_{ab} $ is the vector constraint. We introduce the sympletic variables associated with the Lagrange multipliers through the identifications $\dot{\lambda}^i = A_0^i$, $N^a=\dot{\tilde{\mu}}^a$, $\tilde v_0= \dot{\tilde{\rho}}$, and $ \lambda_{ab}=\dot{\tilde{\beta_{ab}}} $, and we have redefined to $\dot{\tilde{\beta}}_{bd} \equiv \lambda_{bd} - N^a e_{(d\vert{}i\vert{}} F^i_{a\vert{}b)}$. In this manner, the symplectic variables now are given by $\xi^{(1)\alpha} = \left( A^i_a, \; \Pi^a_i, \; e_{ai}, \; \tilde{\lambda}^i, \; \tilde{\mu}^a, \; \tilde{\rho}, \; \tilde{\beta}_{ab} \right)$ and the one form $a^{(1)}_\alpha = \left( \Pi^a_i, \; 0, \; 0, \; \Omega_i, \; \mathcal{H}_a , \; \Omega, \; \phi^{ab}\right)$. With these symplectic variables, we calculate the following symplectic matrix 
\begin{equation}
\label{sym1} 
f^{(1)}_{\alpha\beta}(x,y) =
	 \begin{pmatrix}
	 0 & -\delta^b_a \delta^i_j & 0 & \frac{\delta \Omega_j(y)}{\delta A^i_a(x)} & \frac{\delta \mathcal{H}_b(y)}{\delta A^i_a(x)} & \frac{\delta \Omega(y)}{\delta A^i_a(x)} & 0 \\ 
	 \delta^a_b \delta^j_i  & 0 & 0 & \frac{\delta \Omega_j(y)}{\delta \Pi^a_i(x)} & \frac{\delta \mathcal{H}_b(y)}{\delta \Pi^a_i(x)} & 0 & \frac{\delta \phi^{cd}(y)}{\delta \Pi^a_i(x)} \\ 
	 0 & 0 & 0 & 0 & 0 & \frac{\delta \Omega(y)}{\delta e_{ai}(x)} & \frac{\delta \phi^{cd}(y)}{\delta e_{ai}(x)} \\ 
	 -\frac{\delta \Omega_i(x)}{\delta A^j_b(y)} & -\frac{\delta \Omega_i(x)}{\delta \Pi^b_j(y)} & 0 & 0 & 0 & 0 & 0 \\ 
	 -\frac{\delta \mathcal{H}_a(x)}{\delta A^j_b(y)} & -\frac{\delta \mathcal{H}_a(x)}{\delta \Pi^b_j(y)} & 0 & 0 & 0 & 0 & 0 \\ 
	 -\frac{\delta \Omega(x)}{\delta A^j_b(y)} & 0 & -\frac{\delta \Omega(x)}{\delta e_{bj}(y)} & 0 & 0 & 0 & 0 \\
	  0 & -\frac{\delta \phi^{ab}(x)}{\delta \Pi^d_j(y)} & -\frac{\delta \phi^{ab}(x)}{\delta e_{dj}(y)} & 0 & 0 & 0 & 0 
	  \end{pmatrix}\delta^3(x-y),
\end{equation}
where 
\begin{eqnarray}
\frac{\delta \Omega_k(y)}{\delta A^i_a(x)} &=& \epsilon_{ki}{}^l \Pi^a_l(y), \nonumber\\
\frac{\delta \Omega_k(y)}{\delta \Pi^a_i(x)} &=& \delta^i_k \partial^{}_{ya} + \epsilon_{km}{}^i A^m_a(y), \nonumber\\
\frac{\delta \mathcal{H}_d(y)}{\delta A^i_a(x)} &=& \delta^a_d \left( \Pi^g_i(y) \partial^{}_{yg} - \epsilon_{in}{}^k A^n_g(y) \Pi^g_k(y) \right) - \Pi^a_i(y) \partial^{}_{yd} - \epsilon_{im}{}^k A^m_d(y) \Pi^a_k(y), \nonumber\\
\frac{\delta \mathcal{H}_d(y)}{\delta \Pi^a_i(x)} &=& F^i_{da}(y),  \nonumber\\
\frac{\delta \Omega(y)}{\delta A^i_a(x)} &=& 2\eta^{fga} \left( e_{fi}(y) \partial^{}_{yg} + \epsilon_{ki}{}^l e_{fl}(y) A^k_g(y) \right), \nonumber\\
\frac{\delta \Omega(y)}{\delta e_{ai}(x)} &=& \eta^{abc} F^i_{bc}(y), \nonumber\\
\frac{\delta \phi^{cd}(y)}{\delta \Pi^a_i(x)} &=& \frac{1}{2} \left( \delta^c_a e^{di}(y) + \delta^d_a e^{ci}(y) \right) = \delta^{(c}_a e^{d)i}(y), \nonumber\\
\frac{\delta e^{dk}(y)}{\delta e_{ai}(x)} &=& -e^{ak}(y) e^{di}(y),  \nonumber\\
\frac{\delta \phi^{cd}(y)}{\delta e_{ai}(x)} &=& -\frac{1}{2} \left( e^{di} \Pi^c_k + e^{ci} \Pi^d_k \right) e^{ak} + 2\eta^{cga} \epsilon_{ijk} e^j_g e^{dk} - \eta^{cgh} \epsilon_{klm} e^l_g e^m_h e^{di} e^{ak},  \nonumber\\
\frac{\delta \phi^{cd}(y)}{\delta e_{ai}(x)} &=& -e^{(d\vert{}i} e^{a\vert{}k} \Pi^{c)}_k + 2 e \left( q^{cd} e^{ai} - q^{ca} e^{di} - q^{da} e^{ci} \right).
\end{eqnarray}
The matrix \eqref{sym1} is singular with nullity $=8$,  this means that there are  eight  null vectors.  After long calculations, we find that the null vectors  are given by  the following six
\begin{eqnarray}
\mathcal{V}^{(1)1}_{(k)\alpha} &=& \left( -D_a \delta^i_k, \; \epsilon_{ki}{}^l \Pi^a_l, \; \epsilon_{ki}{}^l e_{al}, \; \delta^i_k, \; 0, \; 0, \; 0 \right), \nonumber \\
\mathcal{V}^{(1)2}_{(c)\alpha} &=& \left( -F^i_{ca}, \; \frac{\delta \mathcal{H}_c}{\delta A^i_a}, \; w^{(c)}_{ai}, \; 0, \; \delta^a_c, \; 0, \; 0 \right),
\end{eqnarray}
where $k, c= 1, 2, 3$. The first null vector is the generator of rotations, and the second is the generator of diffeomorphisms; also, there are  two trivial null vectors that give algebraic relations. Because there are $8$ null vectors to obtain the symplectic tensor, we need to fix gauge conditions. We fix the temporal gauge: three conditions for $A^i_0 = 0$, three for $N^a = 0$, and two for the zero algebraic modes of the triad, say $\vartheta_r(e) = 0$ ($r = 1,2$). In this manner, the symplectic Lagrangian will take the form 
\begin{equation}
\mathcal{L}^{(2)} = \Pi^a_i \dot{A}^i_a + \dot{\tilde{\lambda}}^i \Omega_i + \dot{\tilde{\mu}}^a \mathcal{H}_a + \dot{\tilde{\rho}} \, \Omega + \dot{\tilde{\beta}}_{ab} \phi^{ab} + \dot{\varsigma}^r \vartheta_r(e) + \dot{\chi}_i \tilde{\lambda}^i + \dot{\xi}_a \tilde{\mu}^a, 
\end{equation}
and the symplectic variables are given by  $\xi^{(2)\alpha} = \left( A^i_a, \; \Pi^a_i, \; e_{ai}, \; \tilde{\lambda}^i, \; \tilde{\mu}^a, \; \tilde{\rho}, \; \tilde{\beta}_{ab}, \; \varsigma^r, \; \chi_i, \; \xi_a \right)$, and the one-forms $a^{(2)}_\alpha = \left( \Pi^a_i, \; 0, \; 0, \; \Omega_i, \; \mathcal{H}_a, \; \Omega, \; \phi^{ab}, \; \vartheta_r, \; \tilde{\lambda}^i, \; \tilde{\mu}^a \right)$. Now the symplectic matrix is given by
\begin{equation}
f^{(2)}_{\alpha\beta}(x,y) = \begin{pmatrix}
	 0 & -\delta^b_a \delta^i_j & 0 & \frac{\delta \Omega_j}{\delta A^i_a} & \frac{\delta \mathcal{H}_b}{\delta A^i_a} & \frac{\delta \Omega}{\delta A^i_a} & 0 & 0 & 0 & 0 \\ \delta^a_b \delta^j_i & 0 & 0 & \frac{\delta \Omega_j}{\delta \Pi^a_i} & \frac{\delta \mathcal{H}_b}{\delta \Pi^a_i} & 0 & \frac{\delta \phi^{cd}}{\delta \Pi^a_i} & 0 & 0 & 0 \\ 
	 0 & 0 & 0 & 0 & 0 & \frac{\delta \Omega}{\delta e_{ai}} & \frac{\delta \phi^{cd}}{\delta e_{ai}} & \frac{\delta \vartheta_s}{\delta e_{ai}} & 0 & 0 \\ 
	 -\frac{\delta \Omega_i}{\delta A^j_b} & -\frac{\delta \Omega_i}{\delta \Pi^b_j} & 0 & 0 & 0 & 0 & 0 & 0 & \delta^j_i & 0 \\ 
	 -\frac{\delta \mathcal{H}_a}{\delta A^j_b} & -\frac{\delta \mathcal{H}_a}{\delta \Pi^b_j} & 0 & 0 & 0 & 0 & 0 & 0 & 0 & \delta^b_a \\ 
	 -\frac{\delta \Omega}{\delta A^j_b} & 0 & -\frac{\delta \Omega}{\delta e_{bj}} & 0 & 0 & 0 & 0 & 0 & 0 & 0 \\ 
	 0 & -\frac{\delta \phi^{ab}}{\delta \Pi^d_j} & -\frac{\delta \phi^{ab}}{\delta e_{dj}} & 0 & 0 & 0 & 0 & 0 & 0 & 0 \\
	  0 & 0 & -\frac{\delta \vartheta_r}{\delta e_{bj}} & 0 & 0 & 0 & 0 & 0 & 0 & 0 \\ 
	  0 & 0 & 0 & -\delta^i_j & 0 & 0 & 0 & 0 & 0 & 0 \\ 
	  0 & 0 & 0 & 0 & -\delta^a_b & 0 & 0 & 0 & 0 & 0 
	  \end{pmatrix} \delta^3(x-y),
\end{equation}
we can observe that in the symplectic matrix, $\frac{\delta \vartheta_r}{\delta e_{bj}}$ occurs; however, we do not need to know this variation in explicit  form because it is decoupled from the other entries and the matrix can be inverted. \\
To invert $f^{(2)}_{\alpha\beta}$, we note that the triad variations form a $9 \times 9$ invertible Jacobian matrix $M^I{}_{bj} \equiv \left( \frac{\delta \Omega}{\delta e_{bj}}, \; \frac{\delta \phi^{cd}}{\delta e_{bj}}, \; \frac{\delta \vartheta_r}{\delta e_{bj}} \right)^T,$ where $I = (\Omega, cd, r)$ labels the $1 + 6 + 2 = 9$ algebraic conditions. Its inverse $(M^{-1})^{bj}{}_I$ splits into the three right-inverse sub-blocks $(M^{-1})^{bj}{}_{\Omega}$, $(M^{-1})^{bj}{}_{cd}$, and $(M^{-1})^{bj}{}_{r}$ satisfying $\frac{\delta \Omega}{\delta e_{bj}}(M^{-1})^{bj}{}_{\Omega} = 1$, $\frac{\delta \phi^{ab}}{\delta e_{bj}}(M^{-1})^{bj}{}_{cd} = \delta^{(a}_c \delta^{b)}_d$, $\frac{\delta \vartheta_r}{\delta e_{bj}}(M^{-1})^{bj}{}_{s} = \delta_{rs}$, while all off-diagonal contractions vanish. Thus, by using the relation $\int d^3z \, f^{(2)}_{\alpha\gamma}(x,z) (f^{(2)})^{-1\gamma\beta}(z,y) = \delta^\beta_\alpha \delta^3(x-y)$, the inverse symplectic matrix is explicitly given by
%%%%%
\begingroup
\setlength{\arraycolsep}{4pt}
\renewcommand{\arraystretch}{1.1}
\newsavebox{\BHKmatrixFirst}
\newsavebox{\BHKmatrixSecond}
\newlength{\BHKmatrixWidth}
\sbox{\BHKmatrixFirst}{$\displaystyle
	\left(f^{(2)}\right)^{-1\alpha\beta}(x,y) =
	\left(\begin{array}{cccccc}
		0 & \delta^a_b \delta^j_i & -\dfrac{\delta \phi^{cd}}{\delta \Pi^a_i} (M^{-1})^{bj}{}_{cd} & 0 & 0 & 0 \\[1.5ex]
		-\delta^b_a \delta^i_j & 0 & \dfrac{\delta \Omega}{\delta A^i_a} (M^{-1})^{bj}{}_{\Omega} & 0 & 0 & 0 \\[1.5ex]
		\dfrac{\delta \phi^{cd}}{\delta \Pi^b_j}(M^{-1})^{ai}{}_{cd} & -\dfrac{\delta \Omega}{\delta A^j_b}(M^{-1})^{ai}{}_{\Omega}& 0 & 0 & 0 & -(M^{-1})^{ai}{}_{\Omega}\\[1.5ex]
		0 & 0 & 0 & 0 & 0 & 0 \\[1.5ex]
		0 & 0 & 0 & 0 & 0 & 0 \\[1.5ex]
		0 & 0 &  (M^{-1})^{bj}{}_{\Omega} & 0 & 0 & 0 \\[1.5ex]
		0 & 0 & (M^{-1})^{dj}{}_{ab} & 0 & 0 & (M^{-1})^{di}{}_{ab}\{\Omega,\phi^{ab}\} (M^{-1})^{bj}{}_{\Omega} \\[1.5ex]
		0 & 0 & (M^{-1})^{bj}{}_{r} & 0 & 0 & 0 \\[1.5ex]
		-\dfrac{\delta \Omega_i}{\delta \Pi^b_j} & \dfrac{\delta \Omega_i}{\delta A^j_b} & - (M^{-1})^{bj}{}_{cd}\{\phi^{cd},\Omega_i\} & \delta^j_i & 0 &  (M^{-1})^{bj}{}_{\Omega}\{\Omega,\Omega_i\} \\[1.5ex]
		-\dfrac{\delta \mathcal{H}_a}{\delta \Pi^b_j} & \dfrac{\delta \mathcal{H}_a}{\delta A^j_b} & - (M^{-1})^{bj}{}_{cd}\{\phi^{cd},\mathcal{H}_a\} & 0 & \delta^b_a &  (M^{-1})^{bj}{}_{\Omega}\{\Omega,\mathcal{H}_a\} \\[1.5ex]
	\end{array}\right.\quad\cdots
	$}
\sbox{\BHKmatrixSecond}{$\displaystyle
	\cdots\quad\left.\begin{array}{cccc}
		0 & 0 & \dfrac{\delta \Omega_j}{\delta \Pi^a_i} & \dfrac{\delta \mathcal{H}_b}{\delta \Pi^a_i} \\[1.5ex]
		0 & 0 & -\dfrac{\delta \Omega_j}{\delta A^i_a} & -\dfrac{\delta \mathcal{H}_b}{\delta A^i_a} \\[1.5ex]
		-(M^{-1})^{ai}{}_{cd} & - (M^{-1})^{ai}{}_{s}& (M^{-1})^{ai}{}_{cd} \{\phi^{cd},\Omega_j\} & (M^{-1})^{ai}{}_{cd}\{\phi^{cd},\mathcal{H}_b\} \\[1.5ex]
		0 & 0 & -\delta^i_j & 0 \\[1.5ex]
		0 & 0 & 0 & -\delta^a_b \\[1.5ex]
		- (M^{-1})^{ai}{}_{\Omega}\{\Omega,\phi^{cd}\} (M^{-1})^{bj}{}_{cd} & 0 & - (M^{-1})^{ai}{}_{\Omega}\{\Omega,\Omega_j\} & - (M^{-1})^{ai}{}_{\Omega}\{\Omega,\mathcal{H}_b\} \\[1.5ex]
		0 & 0 & -(M^{-1})^{di}{}_{ab} \{\phi^{ab},\Omega_j\} & -(M^{-1})^{di}{}_{ab} \{\phi^{ab},\mathcal{H}_b\} \\[1.5ex]
		0 & 0 & 0 & 0 \\[1.5ex]
		 (M^{-1})^{bj}{}_{cd}\{\phi^{cd},\Omega_i\} & 0 & \{\Omega_i, \Omega_j\} & \{\Omega_i, \mathcal{H}_b\} \\[1.5ex]
		 (M^{-1})^{bj}{}_{cd}\{\phi^{cd},\mathcal{H}_a\} & 0 & \{\mathcal{H}_a, \Omega_j\} & \{\mathcal{H}_a, \mathcal{H}_b\} \\[1.5ex]
	\end{array}\right)\,\delta^3(x-y),
	$}
\setlength{\BHKmatrixWidth}{\wd\BHKmatrixFirst}
\ifdim\wd\BHKmatrixSecond>\BHKmatrixWidth
\setlength{\BHKmatrixWidth}{\wd\BHKmatrixSecond}
\fi
\begin{gather}
	\resizebox{0.94\textwidth}{!}{%
		\makebox[\BHKmatrixWidth][c]{\usebox{\BHKmatrixFirst}}%
	}\notag\displaybreak[0]\\[3ex]
	\resizebox{0.94\textwidth}{!}{%
		\makebox[\BHKmatrixWidth][c]{\usebox{\BHKmatrixSecond}}%
	}\label{eq:inversa-simplectica}
\end{gather}
\endgroup
%%%%%
where  $\{F, G\} \equiv \frac{\delta F}{\delta A^k_g}\frac{\delta G}{\delta \Pi^g_k} - \frac{\delta F}{\delta \Pi^g_k}\frac{\delta G}{\delta A^k_g}$.
From the inverse symplectic matrix, by means of the Faddeev-Jackiw relation $\{\xi^{(2)\alpha}(x), \xi^{(2)\beta}(y)\}_{FJ} = (f^{(2)})^{-1\alpha\beta}(x,y)$, we readily identify the non-vanishing fundamental brackets for the dynamical phase space variables
\begin{equation}
\{A^i_a(x), \Pi^b_j(y)\}_{FJ} = \delta^b_a \delta^i_j \delta^3(x-y), \qquad \{A^i_a(x), A^j_b(y)\}_{FJ} = 0, \qquad \{\Pi^a_i(x), \Pi^b_j(y)\}_{FJ} = 0.
\end{equation}
Furthermore, the triad $e_{ai}$ is completely determined as an auxiliary field through the algebraic constraints $(\Omega, \phi^{ab}, \vartheta_r)$, while the canonical phase space consists of the $18$ variables $(A^i_a, \Pi^a_i)$ subject to $6$  constraints $(\Omega_i, \mathcal{H}_a)$ and their $6$ gauge-fixing conditions. Consequently, the number of physical degrees of freedom is $DOF = \frac{1}{2}(18 - 2 \times 6) = 3$, which matches the physical degrees of freedom of Husain-Kucha\v{r} gravity.
At this point, it is worth emphasizing why the spatial components $v_a$ cannot be set to zero ($v_a = 0$) by hand as a gauge-fixing condition in action (1), as assumed in \cite{Barbero1994}. Our symplectic analysis reveals two fundamental obstructions. First, as shown by the identity $v_d = \frac{1}{4}\eta_{dga}e^{ig}\Pi^a_i$, the spatial vector $v_d$ is not an arbitrary Lagrange multiplier but parametrizes the three antisymmetric components of the canonical momentum $\Pi^a_i$ in the triad frame. Setting $v_d = 0$ by hand imposes three ad hoc constraints $\eta_{dga}e^{ig}\Pi^a_i = 0$ on the dynamical momenta, which do not arise from any null vector of the symplectic structure.
Second, the role of the scalar constraint $\Omega = 0$ must be taken into account. Since $v_a$ enters the constraint $\Omega^i = \eta^{abc}(v_a F^i_{bc} + \epsilon^i{}_{jk}e^j_a F^k_{bc}) = 0$ without time derivatives, the role of $\Omega = 0$ in action (1) is to fix the auxiliary triad/vector sector rather than constrain the canonical pair $(A^i_a, \Pi^a_i)$. Forcing $v_a = 0$ a priori removes $v_a$ from $\Omega^i$ while retaining $\Omega = 0$ as an independent constraint on $(A^i_a, \Pi^a_i)$, artificially reducing the physical degrees of freedom from $3$ to $2$.%%%%%%%%%%%%%%%%%%%%%%%%%%%%%%%%%%
%%%%%%%%%%%%%%%%%%%%%%%%%%
\section{Alternative action for canonical gravity }
As we demonstrated in the previous section, the condition $v_a = 0$ cannot be imposed as an arbitrary gauge choice in the action (1); instead, if one wishes to recover the $2$ physical degrees of freedom of General Relativity, the vanishing of $v_a$ must emerge dynamically from the variational principle itself. With this motivation, we now analyze the modified action
\begin{equation}
S = \int d^4x \left[ \epsilon^{\alpha\beta\mu\nu} v_\alpha e_{\beta i} F^i_{\mu\nu} - \frac{1}{2} \epsilon^{\alpha\beta\mu\nu} \epsilon_{ijk} e^i_\alpha e^j_\beta F^k_{\mu\nu} + \Lambda^{\alpha\beta} v_\alpha \partial_\beta \phi \right], 
\end{equation}
where we have coupled the vector $v_\alpha$ to a scalar field and  $\Lambda^{\alpha\beta}= - \Lambda^{\beta \alpha}$. In this manner, from the action we identify the symplectic Lagrangian 
\begin{equation}
\mathcal{L}^{(0)} = \Pi^a_i \dot{A}^i_a - \Lambda^{0a} v_a \dot{\phi} - V^{(0)}, 
\end{equation}
where 
\begin{equation}
V^{(0)} = -A^i_0 D_a \Pi^a_i - v_0 \left( \eta^{abc} e_{ai}(\Pi, v) F^i_{bc} + \Lambda^{0a} \partial_a \phi \right) + \eta^{abc} e_{0i} \left( v_a F^i_{bc} + \epsilon^i{}_{jk} e^j_a(\Pi, v) F^k_{bc} \right) - \frac{1}{2}\Lambda^{ab}\left(v_a \partial_b \phi - v_b \partial_a \phi\right),
\end{equation}
we recall that the triad is function of $e_{ai}(\Pi, v)$. Now, from the symplectic Lagrangian, we identify the following symplectic variables 
$\xi^{(0)\alpha} = \left( A^i_0, \; A^i_a, \; \Pi^a_i, \; \phi, \; \Lambda^{0a}, \; \Lambda^{ab}, \; e_{0i}, \; v_0, \; v_a \right)$, and the one-forms $a^{(0)}_\alpha = \left( 0, \; \Pi^a_i, \; 0, \; -\Lambda^{0b} v_b, \; 0, \; 0, \; 0, \; 0, \; 0 \right)$, thus the symplectic matrix will be given by 
\begin{equation}
f^{(0)}_{\alpha\beta}(x,y) = 
\begin{pmatrix}  
0 & 0 & 0 & 0 & 0 & 0 & 0 & 0 & 0 \\ 
0 & 0 & -\delta^a_b \delta^j_i & 0 & 0 & 0 & 0 & 0 & 0 \\ 
0 & \delta^b_a \delta^i_j & 0 & 0 & 0 & 0 & 0 & 0 & 0 \\ 
0 & 0 & 0 & 0 & v_b & 0 & 0 & 0 & \Lambda^{0b} \\ 
0 & 0 & 0 & -v_a & 0 & 0 & 0 & 0 & 0 \\ 
0 & 0 & 0 & 0 & 0 & 0 & 0 & 0 & 0 \\ 
0 & 0 & 0 & 0 & 0 & 0 & 0 & 0 & 0 \\ 
0 & 0 & 0 & 0 & 0 & 0 & 0 & 0 & 0 \\ 
0 & 0 & 0 & -\Lambda^{0a} & 0 & 0 & 0 & 0 & 0  
\end{pmatrix} \delta^3(x-y), 
\end{equation}
from the symplectic matrix, we can observe that there are the following null vectors
\begin{eqnarray}
\mathcal{V}^{(0)1} &=& (\delta^i_k, 0, 0, 0, 0, 0, 0, 0, 0), \nonumber \\
\mathcal{V}^{(0)2}&=& (0, 0, 0, 0, 0, \delta^{[a}_c \delta^{b]}_d, 0, 0, 0),  \nonumber \\
\mathcal{V}^{(0)3} &=& (0, 0, 0, 0, 0, 0, \delta^k_i, 0, 0), \nonumber \\
\mathcal{V}^{(0)4} &=& (0, 0, 0, 0, 0, 0, 0, 1, 0), 
\end{eqnarray}
from these  null vectors, we identify the constraints
\begin{eqnarray}
\label{rel1}
\Omega_i &:& D_a \Pi^a_i = 0, \nonumber \\
\Omega_{ab} &:& v_a \partial_b \phi - v_b \partial_a \phi = 0, \nonumber \\
\Omega^i &:& \eta^{abc}\left( v_a F^i_{bc} + \epsilon^i{}_{jk} e^j_a(\Pi, v) F^k_{bc} \right) = 0, \nonumber \\
\tilde{\Omega} &:& \eta^{abc} e_{ai}(\Pi, v) F^i_{bc} + \Lambda^{0a} \partial_a \phi = 0.
\end{eqnarray}
On the other hand, we will rewrite the constraints in a new fashion. In fact, by using  $B^{ai} \equiv \frac{1}{2} \eta^{abc} F^i_{bc}$ and the constraints  (\ref{rel1}) we obtain 
\begin{equation}
e_{di} \Omega^i - v_d \tilde{\Omega} + \Lambda^{0a} \Omega_{da} = \left( \Pi^c_k F^k_{dc} -\Lambda^{0b}v_d \partial_d \phi \right) \equiv  \tilde{\mathcal{H}}_d,
\end{equation}
where $\tilde{\mathcal{H}}_a(A, \Pi, \phi, \Lambda^{0a}) \equiv \Pi^b_i F^i_{ab} - \Lambda^{0b}v_b \partial_a \phi = 0$ is the diffeomorphism constraint. Furthermore, by introducing $e_{0i} \equiv - N^a e_{ai}$ and  $\tilde{v}_0 \equiv v_0 +  N^a v_a$ and $\tilde{\Lambda}^{ab} \equiv \frac{1}{2}\Lambda^{ab} - \frac{1}{2} N^{[a}\Lambda^{\vert{}0\vert{}b]}$ the symplectic potential can be written as 
\begin{equation}
V^{(0)} = -A^i_0 \Omega_i - N^a \tilde{\mathcal{H}}_a - \tilde{v}_0 \tilde{\Omega} - \tilde{\Lambda}^{ab} \Omega_{ab},
\end{equation}
where we can observe a nice constraint structure. With that structure, we add to the symplectic Lagrangian the constraints with the following Lagrange multipliers: $A^i_0 \to \dot{\lambda}^i$, $N^a \to \dot{\mu}^a$, $\tilde{v}_0 \to \dot{\rho}$, $\tilde{\Lambda}^{ab} \to \dot{\omega}^{ab}$, and we obtain 
\begin{equation}
\label{L1}
\mathcal{L}^{(1)} = \Pi^a_i \dot{A}^i_a -\Lambda^{0b}v_b \dot{\phi} + \dot{\lambda}^i \Omega_i(A, \Pi) + \dot{\mu}^a \tilde{\mathcal{H}}_a(A, \Pi, \phi, \Lambda^{0a}) + \dot{\rho} \tilde{\Omega}(A, \Pi, v, \phi, \Lambda^{0a}) + \dot{\omega}^{ab} \Omega_{ab}(v, \phi) - V^{(1)},
\end{equation}
where $V^{(1)} = \left. V^{(0)} \right\vert{}_{\Omega_i = \tilde{\mathcal{H}}_a = \tilde{\Omega} = \Omega_{ab} = 0} = 0$. Now, from the symplectic Lagrangian  (\ref{L1}) we identify the following symplectic variables $\xi^{(1)\alpha} = \left( A^i_a, \; \Pi^a_i, \; \phi, \Lambda^{0a} , \; v_a, \; \lambda^i, \; \mu^a, \; \rho, \; \omega^{ab} \right)$ and the one-forms $a^{(1)}_\alpha = \left( \Pi^a_i, \; 0, \; -\Lambda^{0a}v_a, 0,  \; 0,  \; \Omega_i, \; \tilde{\mathcal{H}}_a, \; \tilde{\Omega}, \; \Omega_{ab} \right)$, thus,  the  symplectic matrix is given by 
\begin{equation}
f^{(1)}_{\alpha\beta}(x,y) =
 \begin{pmatrix} 
 	0 & -\delta^a_b \delta^j_i & 0 & 0 & 0 & \frac{\delta \Omega_j}{\delta A^i_a} & \frac{\delta \tilde{\mathcal{H}}_b}{\delta A^i_a} & \frac{\delta \tilde{\Omega}}{\delta A^i_a} & 0 \\ 
 	\delta^b_a \delta^i_j & 0 & 0 & 0 & 0 & \frac{\delta \Omega_j}{\delta \Pi^a_i} & \frac{\delta \tilde{\mathcal{H}}_b}{\delta \Pi^a_i} & \frac{\delta \tilde{\Omega}}{\delta \Pi^a_i} & 0 \\ 
 	0 & 0 & 0 & v_b & \Lambda^{0b} & 0 & \frac{\delta \tilde{\mathcal{H}}_b}{\delta \phi} & \frac{\delta \tilde{\Omega}}{\delta \phi} & \frac{\delta \Omega_{cd}}{\delta \phi} \\ 
 	0 & 0 & -v_a & 0 & 0 & 0 & \frac{\delta \tilde{\mathcal{H}}_b}{\delta \Lambda^{0a}} & \frac{\delta \tilde{\Omega}}{\delta \Lambda^{0a}} & 0 \\ 
 	0 & 0 & -\Lambda^{0a} & 0 & 0 & 0 & \frac{\delta \tilde{\mathcal{H}}_b}{\delta v_a} & \frac{\delta \tilde{\Omega}}{\delta v_a} & \frac{\delta \Omega_{cd}}{\delta v_a} \\
 	-\frac{\delta \Omega_i}{\delta A^j_b} & -\frac{\delta \Omega_i}{\delta \Pi^b_j} & 0 & 0 & 0 & 0 & 0 & 0 & 0 \\ 
 	-\frac{\delta \tilde{\mathcal{H}}_a}{\delta A^j_b} & -\frac{\delta \tilde{\mathcal{H}}_a}{\delta \Pi^b_j} & -\frac{\delta \tilde{\mathcal{H}}_a}{\delta \phi} & -\frac{\delta \tilde{\mathcal{H}}_a}{\delta \Lambda^{0b}} & -\frac{\delta \tilde{\mathcal{H}}_a}{\delta v_b} & 0 & 0 & 0 & 0 \\ 
 	-\frac{\delta \tilde{\Omega}}{\delta A^j_b} & -\frac{\delta \tilde{\Omega}}{\delta \Pi^b_j} & -\frac{\delta \tilde{\Omega}}{\delta \phi} & -\frac{\delta \tilde{\Omega}}{\delta \Lambda^{0b}} & -\frac{\delta \tilde{\Omega}}{\delta v_b} & 0 & 0 & 0 & 0 \\ 
 	0 & 0 & -\frac{\delta \Omega_{ab}}{\delta \phi} & 0 & -\frac{\delta \Omega_{ab}}{\delta v_b} & 0 & 0 & 0 & 0 
 	\end{pmatrix} \delta^3(x-y), 
\end{equation}
where 
\begin{eqnarray}
\displaystyle \frac{\delta \Omega_k(y)}{\delta A^i_a(x)} &=& \epsilon{_{ki}}^l \Pi^a_l(y), \nonumber \\
\displaystyle \frac{\delta \Omega_k(y)}{\delta \Pi^a_i(x)} &=& \delta^i_k\partial{_{ya}} + \epsilon{_{km}}^i A^m_a(y),    \nonumber \\
\displaystyle \frac{\delta \tilde{\mathcal{H}}_d(y)}{\delta A^i _a(x)} &=&  \Pi^a_i(y) \partial_{y d} - \epsilon{_{im}}^k A^m_d(y) \Pi^a_k(y)- \delta^{a}_{d} \left[ \Pi^g_i \partial_{yg} - \epsilon_{in}^{k} A^n_g(y) \Pi^g_k(y)\right], \nonumber \\
\displaystyle \frac{\delta \tilde{\mathcal{H}}_d(y)}{\delta \Pi^a_i(x)} &=& F^i_{da}(y),  \nonumber \\
\displaystyle \frac{\delta \tilde{\mathcal{H}}_d(y)}{\delta \phi(x)} &=& - \Lambda^{0b} v_b(y) \partial_{yd},  \nonumber \\
\frac{\delta \tilde{\mathcal{H}}_d(y)}{\delta \Lambda^{0a}(x)} &=& -v_a(y) \partial_{yd} \phi(y), \nonumber  \\
 \frac{\delta \tilde{\mathcal{H}}_d(y)}{\delta v_a(x)} &=& -\Lambda^{0a}(y) \partial_{yd}  \phi(y), \nonumber \\
\displaystyle \frac{\delta \tilde{\Omega}(y)}{\delta A^i_a(x)} &=& 2 \eta^{fga} e_{fi}(\Pi, v)(y)\partial_{yg} + 2 \eta^{fga}e_{fk}(\Pi, v)(y) \epsilon{^k}_{li} A^l_g,  \nonumber \\
\displaystyle \frac{\delta \tilde{\Omega}(y)}{\delta \Pi^a_i(x)} &=& \eta^{fgh} \frac{\partial e_{fk}(\Pi, v)(y)}{\partial \Pi^a_i(x)} F^k_{gh}(y), \nonumber \\
\displaystyle \frac{\delta \tilde{\Omega}(y)}{\delta v_d(x)} &=& \eta^{fgh} \frac{\partial e_{fk}}{\partial v_d}(y) F^k_{gh}(y),  \nonumber \\
\displaystyle \frac{\delta \tilde{\Omega}(y)}{\delta \phi(x)} &=& \Lambda^{0d}(y) \partial_{yd},  \nonumber \\
\displaystyle \frac{\delta \tilde{\Omega}(y)}{\delta \Lambda^{0a}(x)} &=&  \partial_a \phi(y),  \nonumber \\
\displaystyle \frac{\delta \Omega_{cd}(y)}{\delta \phi(x)} &=& v_c(y) \partial_{yd} - v_d(y) \partial_{yc} = 2 v_{[c}(y) \partial^{(y)}_{d]},  \nonumber \\
\displaystyle \frac{\delta \Omega_{cd}(y)}{\delta v_a(x)} &=&  \delta^a_c \partial_d \phi(y) - \delta^a_d \partial_c \phi(y) = 2 \delta^a_{[c} \partial_{d]} \phi(y).
\end{eqnarray}
This matrix is singular and we need fixing the gauge for obtaining a symplectic tensor. The gauge that we will use is $A^i_0 = 0$, $N^a = 0$, $\tilde{v}_0 = 0$, $\dot{\phi} = 1$ this implies that $v_a = 0$. Hence the symplectic Lagrangian takes the form 
\begin{equation}
\mathcal{L}^{(2)} = \Pi^a_i \dot{A}^i_a  + \dot{\lambda}^i \Omega_i + \dot{\mu}^a \tilde{\mathcal{H}}_a + \dot{\rho} \tilde{\Omega} + \dot{\chi}_i \lambda^i + \dot{\xi}_a \mu^a + \dot{\zeta} \rho  + \dot{\theta} (\phi - t) + \dot{\gamma}^av_a- V^{(1)}, 
\end{equation}
and the symplectic variables are given by $\xi^{(2)\alpha} = \left( A^i_a, \, \Pi^a_i, \, \phi, \, v_a, \, \lambda^i, \, \mu^a, \, \rho, \, \chi_i, \, \xi_a, \, \zeta, \, \theta, \, \gamma^a \right)$ and the one-forms  $a^{(2)}_\alpha = \left( \Pi^a_i, \, 0, \, 0, \, 0, \, \Omega_i, \, \tilde{\mathcal{H}}_a, \, \tilde{\Omega}, \, \lambda^i, \, \mu^a, \, \rho, \, \phi - t, \, v_a \right)$. By using the symplectic variables, we calculate the symplectic matrix 
\begin{equation}
f^{(2)}_{\alpha\beta}(x,y) = \begin{pmatrix} 0 & -\delta^a_b \delta^j_i & 0 & 0 & \frac{\delta \Omega_j}{\delta A^i_a} & \frac{\delta \mathcal{H}_b}{\delta A^i_a} & \frac{\delta \Omega}{\delta A^i_a} & 0 & 0 & 0 & 0 & 0 \\ \delta^b_a \delta^i_j & 0 & 0 & 0 & \frac{\delta \Omega_j}{\delta \Pi^a_i} & \frac{\delta \mathcal{H}_b}{\delta \Pi^a_i} & \frac{\delta \Omega}{\delta \Pi^a_i} & 0 & 0 & 0 & 0 & 0 \\ 0 & 0 & 0 & 0 & 0 & 0 & \Lambda^{0d}\partial_{yd} & 0 & 0 & 0 & 1 & 0 \\ 0 & 0 & 0 & 0 & 0 & 0 & \mathcal{W}^a & 0 & 0 & 0 & 0 & \delta^a_b \\ -\frac{\delta \Omega_i}{\delta A^j_b} & -\frac{\delta \Omega_i}{\delta \Pi^b_j} & 0 & 0 & 0 & 0 & 0 & \delta^j_i & 0 & 0 & 0 & 0 \\ -\frac{\delta \mathcal{H}_a}{\delta A^j_b} & -\frac{\delta \mathcal{H}_a}{\delta \Pi^b_j} & 0 & 0 & 0 & 0 & 0 & 0 & \delta^b_a & 0 & 0 & 0 \\ -\frac{\delta \Omega}{\delta A^j_b} & -\frac{\delta \Omega}{\delta \Pi^b_j} & -\Lambda^{0d}\partial_{xd} & -\mathcal{W}^b & 0 & 0 & 0 & 0 & 0 & 1 & 0 & 0 \\ 0 & 0 & 0 & 0 & -\delta^i_j & 0 & 0 & 0 & 0 & 0 & 0 & 0 \\ 0 & 0 & 0 & 0 & 0 & -\delta^a_b & 0 & 0 & 0 & 0 & 0 & 0 \\ 0 & 0 & 0 & 0 & 0 & 0 & -1 & 0 & 0 & 0 & 0 & 0 \\ 0 & 0 & -1 & 0 & 0 & 0 & 0 & 0 & 0 & 0 & 0 & 0 \\ 0 & 0 & 0 & -\delta^b_a & 0 & 0 & 0 & 0 & 0 & 0 & 0 & 0 \end{pmatrix} \delta^3(x-y),
\end{equation}
thus, the inverse is given by 
\begin{equation}
\label{matriz5}
(f^{(2)})^{-1\alpha\beta}(x,y) = \begin{pmatrix} 0 & \delta^b_a \delta^i_j & 0 & 0 & 0 & 0 & 0 & \frac{\delta \Omega_j}{\delta \Pi^a_i} & \frac{\delta \mathcal{H}_b}{\delta \Pi^a_i} & \frac{\delta \Omega}{\delta \Pi^a_i} & 0 & 0 \\ -\delta^a_b \delta^j_i & 0 & 0 & 0 & 0 & 0 & 0 & -\frac{\delta \Omega_j}{\delta A^i_a} & -\frac{\delta \mathcal{H}_b}{\delta A^i_a} & -\frac{\delta \Omega}{\delta A^i_a} & 0 & 0 \\ 0 & 0 & 0 & 0 & 0 & 0 & 0 & 0 & 0 & 0 & -1 & 0 \\ 0 & 0 & 0 & 0 & 0 & 0 & 0 & 0 & 0 & 0 & 0 & -\delta^a_b \\ 0 & 0 & 0 & 0 & 0 & 0 & 0 & -\delta^i_j & 0 & 0 & 0 & 0 \\ 0 & 0 & 0 & 0 & 0 & 0 & 0 & 0 & -\delta^a_b & 0 & 0 & 0 \\ 0 & 0 & 0 & 0 & 0 & 0 & 0 & 0 & 0 & -1 & 0 & 0 \\ -\frac{\delta \Omega_i}{\delta \Pi^b_j} & \frac{\delta \Omega_i}{\delta A^j_b} & 0 & 0 & \delta^j_i & 0 & 0 & \{\Omega_i, \Omega_j\} & \{\Omega_i, \mathcal{H}_b\} & \{\Omega_i, \Omega\} & 0 & 0 \\ -\frac{\delta \mathcal{H}_a}{\delta \Pi^b_j} & \frac{\delta \mathcal{H}_a}{\delta A^j_b} & 0 & 0 & 0 & \delta^b_a & 0 & \{\mathcal{H}_a, \Omega_j\} & \{\mathcal{H}_a, \mathcal{H}_b\} & \{\mathcal{H}_a, \Omega\} & 0 & 0 \\ -\frac{\delta \Omega}{\delta \Pi^b_j} & \frac{\delta \Omega}{\delta A^j_b} & 0 & 0 & 0 & 0 & 1 & \{\Omega, \Omega_j\} & \{\Omega, \mathcal{H}_b\} & \{\Omega, \Omega\} & -\Lambda^{0d}\partial_{xd} & -\mathcal{W}^b \\ 0 & 0 & 1 & 0 & 0 & 0 & 0 & 0 & 0 & \Lambda^{0d}\partial_{yd} & 0 & 0 \\ 0 & 0 & 0 & \delta^b_a & 0 & 0 & 0 & 0 & 0 & \mathcal{W}^a & 0 & 0 \end{pmatrix} \delta^3(x-y),
\end{equation}
where $\mathcal{W}^a \equiv \left.\frac{\delta \tilde{\Omega}}{\delta v_a}\right\vert{}_{v_a=0} = \eta^{fgh} F^k_{gh}  \left.\frac{\partial e_{fk}}{\partial v_a}\right\vert{}_{v_a=0}$. From the inverse symplectic matrix \eqref{matriz5}, as in the previous case, we obtain the non-vanishing fundamental Faddeev--Jackiw brackets among the dynamical phase space variables:
  \begin{eqnarray}
  \{A^i_a(x), \Pi^b_j(y)\}_{FJ} = \delta^b_a \delta^i_j \delta^3(x-y), \qquad \{A^i_a(x), A^j_b(y)\}_{FJ} = 0, \qquad \{\Pi^a_i(x), \Pi^b_j(y)\}_{FJ} = 0,
  \end{eqnarray}  
All brackets between the gravitational variables $(A^i_a, \Pi^a_i)$ and the auxiliary sector $(\phi, v_a)$ vanish identically. This confirms that the clock gauge $\dot{\phi} = 1$, which enforces $v_a = 0$ and $\partial_a \phi = 0$, fully decouples the auxiliary vector $v_a$ and reduces the triad $e_{ai}$ to depend only on the canonical momenta $\Pi^a_i$. The resulting phase space consists of 18 canonical variables $(A^i_a, \Pi^a_i)$, subject to seven constraints of General Relativity ($\Omega_i, \mathcal{H}_a, \Omega$) and their seven gauge-fixing conditions. Counting the physical degrees of freedom yields $\text{DOF} = \frac{1}{2}(18 - 2 \times 7) = 2$, which matches the two physical degrees of freedom in Einstein's General Relativity as recovered from the modified Barbero–Husain–Kuchař action.

\section{Conclusions}
In this paper, we carry out a full FJ symplectic analysis of the Barbero-Husain-Kucha\v{r} (BHK) gravity action and propose a consistent dynamical extension which recovers the two physical degrees of freedom of GR.\\
In the first part, we analyzed the original BHK action. We showed that the definition of the canonical momenta $\Pi^a_i \equiv \eta^{abc}(2v_b e_{ci} - \epsilon_{ijk}e^j_b e^k_c)$ mixes the nine components of the triad $e_{ai}$ with the three spatial components of the auxiliary covector $v_a$. By inverting this relation in the triad frame, we found two structural features overlooked in previous treatments: first, the spatial components $v_d$ are not arbitrary Lagrange multipliers but uniquely parametrize the three antisymmetric components of the canonical momenta via the exact identity $v_d = \frac{1}{4}\eta_{dga}e^{ig}\Pi^a_i$; second, eliminating $v_a$ from the symplectic variables leaves six independent symmetric constraints $\phi^{ab}(\Pi, e) = 0$ relating the symmetric part of the momenta to the spatial metric $q^{ab} = e^{ia}e^{ib}$. Considering $\phi^{ab} = 0$ in the symplectic Lagrangian revealed that the symplectic matrix $f^{(1)}_{\alpha\beta}$ has a nullity of $8$, corresponding to three generators of internal $SU(2)$ rotations, three generators of spatial diffeomorphisms, and two algebraic zero modes in the auxiliary triad sector. After imposing the temporal gauge together with two algebraic conditions called  $\vartheta_r(e)=0$ ($r=1,2$), we explicitly inverted the symplectic matrix and extracted the fundamental FJ brackets $\{A^i_a(x), \Pi^b_j(y)\}_{FJ} = \delta^b_a\delta^i_j\delta^3(x-y)$. This analysis proves that the nine components of the triad $e_{ai}$ are completely fixed algebraically by $(\Omega, \phi^{ab}, \vartheta_r)$, leaving the $18$-dimensional canonical phase space $(A^i_a, \Pi^a_i)$ subject only to the $6$ constraints $(\Omega_i, \mathcal{H}_a)$. Consequently, the unaltered BHK action propagates $\text{DOF} = \frac{1}{2}(18 - 2 \times 6) = 3$ physical degrees of freedom, matching standard HK gravity. Furthermore, this shows why setting $v_a = 0$ by hand as a gauge choice is mathematically inconsistent: it imposes three \textit{ad hoc} conditions on a sector that admits only two algebraic gauge symmetries, artificially truncates the antisymmetric part of the momenta $\Pi^a_i$, and illegitimately promotes $\Omega = 0$ from an auxiliary triad condition to a phase-space constraint.\\
We also extend our work further. In Section III we considered an alternative action in which the vector field $v_\alpha$ is coupled to a scalar field $\phi$ via the antisymmetric tensor term $\Lambda^{\alpha\beta}v_\alpha\partial_\beta\phi$ (without breaking the symplectic structure). Applying the Faddeev-Jackiw formalism to this extended action, we find that the new coupling leads to the constraint $\Omega_{ab} = v_a\partial_b\phi - v_b\partial_a\phi = 0$, together with modified diffeomorphism and scalar constraints. Importantly, if we fix the relational clock gauge $\dot{\phi} = 1$ ($\phi = t$, $\partial_a\phi = 0$) and also impose the temporal gauge conditions, the vanishing of the spatial vector $v_a = 0$ arises dynamically from the variational principle rather than being imposed by hand. The inverse symplectic matrix (38) confirms that the auxiliary sector $(\phi, \Lambda^{0a}, v_a)$ decouples completely from the gravitational variables $(A^i_a, \Pi^a_i)$ and the canonical Faddeev-Jackiw brackets are preserved in the reduced sector. In the reduced sector, the triad is a function of the canonical momenta $\Pi^a_i$ alone and the 18-dimensional phase space is governed by seven constraints $(\Omega_i, \mathcal{H}_a, \Omega)$ and their seven gauge-fixing conditions, giving $\text{DOF} = \frac{1}{2}(18 - 2 \times 7) = 2$ physical degrees of freedom.\\
Finally, our results demonstrate the efficiency and geometric transparency of the Faddeev-Jackiw symplectic formalism for first-order gravitational actions with complicated auxiliary sectors. By avoiding the manipulation of primary momenta that would otherwise be associated with the non-dynamical multipliers, the zero mode structure is in a position to directly identify inconsistent gauge truncations and to guide the construction of consistent dynamical modifications. The next step will be to fully quantize the modified action in the relational clock gauge and explore its extension to two connection formulations of Lorentzian GR.

\end{document}